\documentclass[%
 reprint,
 amsmath,amssymb,
 aps,
]{revtex4-2}

\usepackage{amsthm}
\usepackage{graphicx}% Include figure files
\usepackage{dcolumn}% Align table columns on decimal point
\usepackage{bm}% bold math

\begin{document}

\preprint{APS/123-QI}

\title{Bi-Quadratic Improvement in Conditional Quantum Search}

\author{Akankshya Dash*}
\affiliation{%
 School of Applied Sciences, Kalinga Institute of Industrial Technology (KIIT) University,\\
 Bhubaneswar, 751024, Odisha, India
}%
 
%\author{S}%
 \email{akankshya1289@gmail.com}
%\affiliation{%}%

%\collaboration{%}%\noaffiliation

\author{Biswaranjan Panda}
 
\email{biswaranjanpanda2002@gmail.com}
 
\affiliation{Indian Institute of Science Education and Research (IISER),\\
Berhampur, 760005 Odisha, India}%
%\affiliation{
%}%

\author{Arun K Pati}

\email{akpati@iiit.ac.in}

\affiliation{Center for Quantum Science and Technology (CQST), International Institute of Information Technology (IIIT),\\
Hyderabad, 500032, Telangana}%
%\affiliation{}

%\collaboration{%}%\noaffiliation

\date{\today}% It is always \today, today,

\begin{abstract}

The Grover search algorithm performs an unstructured search of a marked item in a database quadratically faster than classical algorithms and is shown to be optimal. Here, we show that if the search space is divided into two blocks with the local query operators and the global operators satisfy certain condition, then it is possible to achieve an improvement of bi-quadratic speed-up. Furthermore, we investigate the bi-quadratic speed-up in the presence of noise and show that it can tolerate noisy scenario. This may have potential applications for diverse fields, including database searching, and optimization, where efficient search algorithms play a pivotal role in solving complex computational problems.

\end{abstract}


\maketitle

%\tableofcontents

\section{Introduction} \label{into}

Quantum computing has emerged as a transformative paradigm, promising to revolutionize various domains by solving problems that were once considered computationally intractable for classical computers. Quantum computers utilize the unique properties of quantum mechanics to perform computations at an unprecedented speed and efficiency compared to classical computers 
\cite{deu1,deu2,deu3}. However, the technology is still in its infancy, and there remain significant challenges to be addressed, including error correction \cite{1}, scalability \cite{2}, and reliability \cite{3}. Despite of these challenges, the rapid pace of innovation and development in quantum computing suggests that it will undoubtedly revolutionize the way we process and communicate information in the future.

Quantum algorithms have showcased their performance in quantum speed up, big data analysis, optimization, simulation and cryptographic security. Quantum search algorithm \cite{9,20,21,4,5} can be potentially applied for efficiently counting the number of solutions for a given search problem, speedup the solution of NP-complete problems, speed up the search for keys to cryptosystems \cite{6}, finding the shortest path between two cities \cite{7}, and searching problems or extracting statistics in unstructured databases \cite{8} faster than the classical search.

The classical brute-force approach to searching an unsorted database requires examining each item one by one, leading to a linear time complexity. Grover's algorithm \cite{9}, in stark contrast, offers a quadratic speedup, fundamentally altering the landscape of search algorithms. By utilizing quantum superposition and interference, it efficiently identifies the target item from an unstructured database with remarkable speed-up, making it a promising solution to some of the most complex search problems. 

Over the year, there have been several development in the realm of Grover's algorithm. One noteworthy development is the incorporation of structured search \cite{10}, where the overall success rate is contingent upon the multiplication of individual search success rates. This is particularly pivotal in scenarios characterized by relatively modest dimensions, as the conventional Grover's algorithm may exhibit suboptimal performance under such circumstances. Consequently, to address this challenge, a series of adapted search algorithms have been introduced \cite{13,14,15,16,17,18,19}. Also, the role of quantum entanglement in the Grover algorithm has been explored and it was shown that the quantum feature such as entanglement is neccessary for the desired speed-up \cite{20,23}.

In the pursuit of enhancement of algorithmic efficiency, our study delves into the intricate realm of the Grover search algorithm. Our quest involves strategically partitioning of the search space into two discrete blocks, coupled with the imposition of specific condition upon local query operators. The perceptive outcome of our study unveils a noteworthy prospect with an attainment of bi-quadratic speed-up. Moreover, our analysis extends to scrutinize the algorithm's resilience in the achievement of bi-quadratic speed-up in the presence of noise, demonstrating its capacity to tolerate noisy scenarios. Our inquiry is directed towards the examination of the behavior of the ultimate state when subjected to the perturbing effects of a noisy channel, exemplified by the amplitude damping channel \cite{adc,adc1,adc2,adc3}. Considering the intrinsic multilevel nature of our system, we utilize the multi-level amplitude damping channel \cite{24,25,26,27,28,29,30,31,32,33,34,35,36,37,38,39,40}, specifically designed to illuminate the decay of energy levels within higher dimensional of quantum system. These advancements resonate across diverse applications in fields such as database searching and optimization, underscoring the algorithm's pivotal role in addressing complex computational challenges.

This paper is structured as follows: Firstly, in section-\ref{se1}, an overview of the Grover algorithm is presented and concurrently we introduce a resilient variant of the conditional quantum search, featuring a bi-quadratic speed-up. This speed-up is achieved when the search space is partitioned into two blocks, and the local query operators adhere to specific conditions for both pure and mixed states. 
In section-\ref{se2}, we delve into an examination of the bi-quadratic speed-up in the presence of noise. We showcase its capacity to endure and remain effective in a noisy scenario. In section-\ref{se3}, we conclude our findings.

%we assert that a bi-quadratic acceleration can be attained if the local query operators adhere to specific conditions. This speed-up holds promise for diverse fields where efficient search algorithms play a pivotal role in addressing intricate computational problems.

\section{Bi-Quadratic Improvement in Quantum Search}\label{se1}

In search algorithm,  we have a function $f(x)$ which results a value $f(x) =1$, if $x$ is the search item. If the item $x$ 
 is not a solution to the search problem, then $f(x) = 0$. The search problem aims to find find an item $y$ such that $f(y) =1$. In the Grover algorithm  we have a classical function $f(x) : \{0, 1, 2,  \cdots,  N-1 \}  \rightarrow  \{0, 1\} $, where $N$ is the size of the database. 
 Our goal is to find $y$ such that $f(y) = 1$.
 
 The Grover algorithm fundamentally entails the execution of a series of Grover operators, ultimately finding the desired target state in the requisite number of iterations. Each Grover operator corresponds to a subtle rotation within the two-dimensional subspace. 
For $n$ qubit system, we take the initial state as $|000...0\rangle$ and to make it equal amplitude we apply the Hadamard transformation to maintain superposition on every qubit as expressed by
\begin{equation}
    |\Psi_0 \rangle= H^{\otimes n}|000...0\rangle = \frac{1}{\sqrt{N}} \sum_{x=0} ^{N-1} |x\rangle ,
\end{equation}
where $N$ = 2$^n$ and $H^{\otimes n}$ is the Hadamard transformation. After applying the Hadamard transformation, we apply the Grover Operator $G$ as given by
\begin{equation} \label{e1}
    G = -I_0 I_y, 
\end{equation}
where $I_y = (I - 2  | y  \rangle \langle y | ) $ and $I_0 = (I - 2  | \Psi_0  \rangle \langle \Psi_0 | )$.
Geometrically, the Grover operator, $G$ rotates the initial state towards the target state $|y\rangle$.  As we proceed through the $k$th iteration of the Grover operator, we achieve  convergence towards our desired state an approximate order of $O$($\sqrt{N}$) steps. This outcome is particularly remarkable when contrasted with classical search algorithms. The ensuing state is depicted as

\begin{equation}
    |\Psi_k\rangle = \frac{cos\theta_k}{\sqrt{N-1}} \sum_{x \ne y}^{N-1}|x \rangle + sin\theta_k |y\rangle,
\end{equation}
where $\theta_k$ = $(2k + 1)\theta$ and $\sin \theta = \frac{1}{\sqrt N}$. For $k$ = $O(\sqrt{N})$, we reach the target state $|y\rangle$. This gives quadratic speed up for the Grover algorithm.

Since it has previously been established that surpassing the quadratic speedup achieved by Grover's algorithm in a general case is unattainable, one might ponder: What happens if we were to introduce specific conditions in the search item? To answer this question, we find that it is possible to go beyond  $O(\sqrt{N})$ by imposing a particular condition, which is succinctly elucidated in the ensuing sections.

\subsection{Conditional Quantum Search for Pure State }

Here, we will introduce the Conditional Quantum Search (CQS) to enhance the speed up of our search within an unsorted database. This constraint imparts a notable improvement, significantly amplifying the speed-up achieved in our search endeavors. This augmentation assumes a heightened significance, especially when our quest aligns with the prescribed criteria, as it empowers us to harness the capabilities of the conditional quantum search algorithm.

%The stipulated condition, delineated for the applicability of our algorithm, is as follows.
Imagine that the marked item has an inherent structure and we are able to partition our target state into two discernible components.
For example, the telephone directory has a person's name which has the first name and the surname. One can directly search the full name and also search the first name and the surname separately. In the later case, the search space is divided into two blocks and the combination of marked items from both the database give the marked item in the larger database. Under such scenario, we can then effectively apply our tailored searching algorithm. We have provided one example here, but practical instances are abound where this condition can be met.
%and we furnish illustrative examples in subsequent sections of our work.
To expound upon this, in any scenario where our target item is denoted as $ab$, should we possess the capability to disentangle and isolate the constituents $a$ and $b$ within our search problem, the efficacy of our algorithm becomes readily applicable and will be notably advantageous.

%\subsection{Proof: For O(N$^{1/4}$)}

Let us consider a quantum computer with $2n$ qubits. With this quantum resister, we can encode $2^{2n}$ = $N^2$ entries of a database. Let the marked item is denoted as $|Y\rangle$. If we apply the Grover search operator in this context, knowing that it offers quadratic speed-up, we can achieve our target state in just $O$($N$) steps. 
In the sequel, we can achieve a bi-quadratic speed-up if the target state can be decomposed into two separate parts using conditional quantum search (CQS).

%For a pure state, suppose we have $N^2$ entries in 
Now, we divide the $2n$ qubits into two blocks, each with $n$ qubits. Let us imagine that the target item encoded in the state $|Y\rangle = |1 \rangle^{\otimes 2n}$ has an inherent structure, i.e., the target item $|Y\rangle$ is combination of target items for two small blocks. For simplification, let us denote the target item for each block as $|y\rangle$ = $|111.....11\rangle = |1 \rangle^{\otimes n}$ and $|y\rangle$ = $|111.....11\rangle = |1 \rangle^{\otimes n}$ with the condition that $|Y\rangle$ = $|y \otimes y\rangle$. We also equipped with local query operators, $I_y$ = $I-2|y\rangle \langle y|$ in each block as well as the global query, $I_Y$ = $I-2|Y\rangle \langle Y|$. Note that $I_y \otimes I_y$ $\ne I_Y$.
Nevertheless, we have
$$ I_y \otimes I_y |Y\rangle =  -I_Y |Y\rangle ,$$
i.e., $I_Y \otimes I_Y |Y\rangle$ = $I_Y |Y\rangle$ up to a phase.
This symmetry matching condition helps us to obtain a bi-quadratic improvement in the conditional quantum search.

Now, consider the initial state of $2n$ qubits as given by

\begin{equation} \label{eqns4}
    |\Psi_0 \rangle = H^{\otimes n}|0\rangle ^{\otimes n} \otimes H^{\otimes n} |0\rangle^{\otimes n}.
\end{equation}
Next, we apply $G \otimes G$ operator $k$ times on the initial state $|\Psi_0\rangle$ and obtain

\begin{equation}
\begin{aligned}
|\Psi_k \rangle = &\left(\cos\theta_k |\overline{y}\rangle  + \sin\theta_k |y\rangle \right) \\
& \otimes \left(\cos\theta_k |\overline{y}\rangle  + \sin\theta_k |y\rangle \right) ,
\end{aligned} 
\end{equation}
where $|\overline y\rangle=\frac{1}{\sqrt{N-1}}\sum_{x \ne y} |x\rangle$.

At last, we apply the $I_{Y}$ operator, which acts as a reflection operator for the target state $|Y\rangle = |y\rangle \otimes |y\rangle$. Thus, the final state is given by

\begin{align}\label{eq9}
   |\Psi_k\rangle &= I_Y |\Psi_k\rangle  \nonumber \\
   &= ( \cos\theta_k |\overline{y}\rangle + \sin\theta_k |y\rangle ) \otimes ( \cos\theta_k |\overline{y}\rangle + \sin\theta_k |y\rangle ) \nonumber \\
   &\quad - 2\sin^2 \theta_k |Y\rangle\\
   &= cos^2\theta_k |\overline{y}\rangle |\overline{y}\rangle -sin^2\theta_k |Y\rangle + sin\theta_k cos\theta_k (|\overline{y}y\rangle + |y\overline{y}\rangle) .
\end{align}

Here, the application of $I_{Y}$ entangles two blocks. If we measure the target state $|Y\rangle$, we will find the marked item.
From Eq.(\ref{eq9}), we see that to find the measured item $|Y\rangle$, we have used the oracle $I_y$~~ $O$($2\sqrt{N}$) times and $I_Y$ once. Therefore this shows that by application of Grover Oracle $O$($2\sqrt{N}+1$) $\approx$ $O$($\sqrt{N}$) we are able to find marked item. Without the matching condition it will take $O$($N$) steps. Classically, it takes $O$($N^2$) steps. Hence, our approach has the capability to identify designated items with a speed enhancement of $O$($N^{1/4}$).

%Kepp this para where required

\subsection{Conditional Quantum Search For Mixed State}

Here, we will discuss how the conditional quantum search is affected if the initial state is not pure but a mixed state. For the sake of illustration, we consider a pseudo-pure state which is an admixture of pure state and a random state. This kind of state arises naturally in many noisy scenarios. 
Suppose the noise parameter, denoted as $\epsilon$, is introduced into the density matrix of our quantum system. The density matrix of the pseudo-pure state for $2n$ qubits can be expressed as 

\begin{equation}
    \rho_0 = \epsilon|\Psi_0 \rangle \langle\Psi_0| + (1 - \epsilon)\frac{I}{N^2},
\end{equation}
where the initial state is given in Eq. (\ref{eqns4}). Now, we apply the Grover operator $(G \otimes G)$ on the density operator $\rho_0$ $k$ times. After applying the Grover operator to the above density operator, we obtain
\begin{equation}
\rho_k = (G \otimes G)\rho_0 (G \otimes G) .
\end{equation}

At last we apply the global query operator $I_Y$ which entangles the two block of quantum registrars. The final state is given by

\begin{equation}
\begin{aligned}
 \rho_k'= &I_Y (G \otimes G) \rho_0 (G \otimes G)I_Y\\
 &= \epsilon |\Psi_k \rangle \langle \Psi_k| + (1-\epsilon)\frac{I}{N^2} .\\ 
\end{aligned}
\end{equation}

Now we measure our target state $|Y\rangle$. 
%So the measurement operator is defined as follows: $$\Pi_Y = |Y\rangle \langle Y|$$
The probability of finding the target state $|Y\rangle$ can be calculated by applying the measurement operator on $\rho_k'$ and can be expressed as 
$$P(|Y\rangle) = Tr[\Pi_Y (\rho_k')] = Tr[|Y\rangle \langle Y|(\rho_k')] $$ 
%By doing above calculation finally we get the probability of finding our target state $|Y\rangle$, is given as:
\begin{equation}
    = (1-\epsilon) + \epsilon sin^4 \theta_k .
\end{equation}
The above equation expresses the effect of noise at $k$th iteration. For $k= O(\sqrt N)$, the effect of noise cancels and we can find the marked item with unit probability.

\section{Bi-Quadratic Speed-up in Noise Scenario}\label{se2}

 In this section, we delve into the behavior of the final state $|\Psi_k\rangle$ in the presence of a noisy channel, such as the amplitude damping channel (ADC). Given that our system is inherently multi-level, we employ the multi-level amplitude damping (MAD) channel (referenced as \cite{19}), specifically tailored to elucidate the energy level decay within a higher dimensional quantum system. The general basis of this channel is represented as $\{|i\rangle\}$, where $i$ = 0, 1, 2..., $d-1$. An MAD channel \( \mathcal{D} \) is a completely positive trace-preserving mapping that operates on the set 
 \( L( {\cal H}_A) \) of linear operators on the quantum system \( {\cal H}_A \). This channel is characterized by a set of Kraus operators, which define the evolution of the system under the influence of the channel. The MAD channel \( \mathcal{D} \) and Kraus operators are given by

 \begin{equation}
     \mathcal{D}(\rho) = K_0 \rho K_0 ^{\dagger} + \sum_{0 \le i \le j \le d-1} \eta_{ji} |i\rangle\langle i| \langle j|\rho|j\rangle ,
 \end{equation}
 where $ K_0 = |0\rangle \langle 0| + \sum_{1 \le j \le d-1} \sqrt{1 - \kappa_j} |j \rangle \langle j|  $
 and  $K_{ij} = \sqrt{\eta_{ji}} |i\rangle\langle j| $.
 
Here, $\rho$ is the density matrix and $\eta_{ji}$ describes about the decay rate from the j-th to i-th level which follows the conditions: 0 $\le$ $\eta_{ji}$ $\le$ 1, $\forall$ $i, j$ s.t. 0 $\le$ $i$ $<$ $j$ $\le$ $d-1$.  and $\kappa_j$ = $\sum_{0 \le j < j} \eta_{ji}$ $\le$ 1, $\forall j$ = 1, 2, ..., $d-1$. 

If this noisy channel acts on the density matrix $\rho_k$ = $|\Psi_k \rangle\langle\Psi_k|$,  then the output state $\mathcal{D}(\rho_k)$ is given by

\begin{equation}
\begin{aligned}
    \mathcal{D}(\rho_k) = 
    &\left|\langle 0|\Psi_k\rangle\right|^2 |0\rangle \langle 0| 
    + \sum_{1 \le J \le N^2 - 1} \left|\langle J|\Psi_k\rangle\right|^2  |J \rangle \langle J| +\\
    &\sum_{0 \le \overline{Y} < J \le N^2 - 1} \eta_{J\overline{Y}} \left| \langle J|\Psi_k\rangle\right|^2  |\overline{Y}\rangle\langle \overline{Y}| .\\ 
\end{aligned}
\end{equation}

Now, measuring the output state of MAD channel $\mathcal{D}(\rho_k)$ in the target state $|Y\rangle$, gives us 
\begin{equation*}
\begin{aligned}
  \langle Y|\mathcal{D}(\rho_k)|Y\rangle = &\left|\langle 0|\Psi_k\rangle\right|^2 \left|\langle Y|0\rangle \right|^2 +\\
    &\sum_{1 \le J \le N^2 - 1} (1 - \kappa_J) \left|\langle JH|\Psi_k\rangle\right|^2  \left|\langle Y|J \rangle \right|^2 +\\ 
    &\sum_{0 \le \overline{Y} < J \le N^2 - 1} \eta_{J\overline{Y}} \left| \langle J|\Psi_k \rangle\right|^2 \left|\langle Y|\overline{Y}\rangle\right|^2 .\\ 
    \end{aligned}
\end{equation*}

By performing the calculations for all the terms, we ultimately obtain the transformed expression of $\langle Y|\mathcal{D}(\rho_k)|Y\rangle$ as given by

\begin{equation*}
\begin{aligned}
    \langle Y|\mathcal{D}(\rho_k)|Y\rangle =
     &(1 - \kappa_Y) sin^4 \theta_k + \\
    &(\sum_{\overline{Y} \ne Y} \eta_{Y\overline{Y}} cos^2 \theta_k - \eta_{YJ} sin^2 \theta_k + \\
    &2\eta_{yy^{'}\overline{y}y} cos\theta_k sin\theta_k)^2 .\\
    \end{aligned}
\end{equation*}

Finally, invoking the definition of the Grover algorithm,  for $k = O(\sqrt N)$ it becomes evident that $\sin\theta_k = 1$ and all other terms vanish. Consequently, the probability of successfully finding the target item is given by

\begin{equation}\label{eqn1}
    P(\mathcal{D}(\rho_k)) = 1 - (\kappa_Y + \eta_{YJ}) .
\end{equation}
Eq. (\ref{eqn1}) articulates the fact that for $(\kappa_Y + \eta_{YJ}) = 0$, we have the noise-less scenario and the target state can be found with certainty.
When the noise is present, we have $(\kappa_Y + \eta_{YJ})$ is nonzero and the success probability reduces.

\section{Conclusion}\label{se3}

The Grover search algorithm has undeniably been a cornerstone in the evolution of quantum algorithms, providing a quadratic speed-up. This paper has contributed to the discourse by demonstrating that under specific conditions, leveraging the intrinsic structure of marked items and employing local query operators, the algorithm's prowess can be elevated to a bi-quadratic speed-up.  Our exploration into the algorithm's resilience in the face of noise underscores its ability to tolerate a noisy scenario. The implications of our findings reverberate across various domains, particularly in conditional database searching and optimization, reaffirming the algorithm's crucial role in unraveling intricate computational challenges.  As quantum computing advances, our study propels the Grover search algorithm from a catalyst of innovation to a versatile tool poised to shape the landscape of future quantum applications.

\begin{acknowledgments}
AD and BP acknowledge IIIT, Hyderabad for the invaluable support and enriching academic environment during our visit, which significantly enhanced the research presented in this project. 
\end{acknowledgments}

\end{document}